 \documentstyle[preprint,aps,epsfig]{revtex}

\tightenlines

\preprint{\vbox{  
\hbox{IFT-P.040/99}   
\hbox{May 1999} }}  
\begin{document}  
\draft 
\title{
A remark on the muonium to antimuonium conversion\\  in a 331 model}
\author{V. Pleitez} 
\address{
Instituto de F\'\i sica  Te\'orica\\ 
Universidade  Estadual Paulista\\
Rua Pamplona, 145\\ 
01405-900-- S\~ao Paulo, SP\\ 
Brazil } 
\date{\today}
\maketitle 
\begin{abstract}  
Here we analyze the relation between the search for muonium to antimuonium
conversion and the 331 model with doubly charged bileptons. We show that the 
constraint on the mass of the vector bilepton obtained by 
experimental data can be evaded even in
the minimal version of the model since there are other contributions to 
that conversion. We also discuss the condition for which the experimental data 
constraint is valid.
\end{abstract}
\narrowtext   

Recently a new upper limit for the spontaneous transition
of muonium ($M\equiv\mu^+e^-$) to antimuonium ($\overline{M}\equiv
\mu^-e^+$) has been obtained~\cite{exp}. 
This implies constraints upon the models that induce the $M\to\overline{M}$ 
transition.
One of them is the 331 model proposed some years ago~\cite{331}. 
Here we would like to discuss 
the conditions in which this constraint can be evaded even in the context of 
the minimal version of the model (minimal in the sense that no new symmetries 
or fields are introduced). 

In that model in the lepton sector the charged physical mass 
eigenstates (unprimed fields) are related to the weak eigenstates (primed 
fields) through unitary transformations ($E_{L,R}$) as follows:
\begin{equation}
l'_L=E_Ll_L,\quad l'_R=E_Rl_R,
\label{1}
\end{equation}
where $l=e,\mu,\tau$.
It means that the doubly charged vector bilepton, $U^{++}_
\mu$, interacts with the charged leptons through the current given by
\begin{equation}
J^\mu_{U^{++}}=-\frac{g_{3l}}{\sqrt2}\,\overline{l_L^c}\gamma^\mu {\cal K} l_L,
\label{bu}
\end{equation}
where  ${\cal K}$ is the unitary matrix defined as ${\cal K}=E^T_RE_L$ in the 
basis in which the interactions with the $W^+$ are diagonal ($\nu^\prime_L
=E_L\nu_L$)~\cite{liung}.

In the theoretical calculations of the $M\to\overline{M}$ transition induced 
by a doubly charged vector bilepton so far only the case  
${\cal K}={\bf1}$ has been considered~\cite{teo}.
Although this is a valid simplification it does not represent the most general
case in the minimal 331 model. 
In fact, in that model in the quark sector all left-handed mixing 
matrices  survive in different places of the lagrangian density~\cite{dumm}. 
In the lepton sector both, left- and right-handed mixing matrices
survive in the interactions 
with the doubly charged vector bilepton as in 
Eq.~(\ref{bu}) and also with doubly and singly charged scalars (see below).
Hence, these mixing matrices as ${\cal K}$ in Eq.~(\ref{bu}) have the same 
status than the Kobayashi-Maskawa mixing matrix in the
context of the standard model in the sense that they must be determined by
experiment. In Ref.~\cite{exp} it is recognized that their bound is valid
only for the flavor diagonal bilepton gauge boson case {\it i.e.}, 
${\cal K}={\bf1}$.
If nondiagonal interactions, like in Eq.~(\ref{bu}), are assumed the new upper 
limit on the conversion probability in the $M\to\overline{M}$ system implies 
\begin{equation}
M_{U^{++}}>g_{3l}\vert{\cal K}_{\mu\mu}\vert\vert{\cal K}_{ee}\vert\,2.6 
\;{\rm TeV}=850\;
\vert{\cal K}_{\mu\mu}\vert\vert{\cal K}_{ee}\vert\;
{\rm GeV}.
\label{bound}
\end{equation}
If $\vert{\cal K}_{\mu\mu}\vert\vert{\cal K}_{ee}\vert\approx0.70 $ we get a 
lower bound of 600 GeV for the doubly charged bilepton which is compatible 
with the upper bound obtained by theoretical arguments~\cite{phf}.

The following is more important. Besides the contribution of the vector 
bileptons there are also the doubly charged and the neutral scalar ones. 
To consider only the vector bileptons is also a valid approximation since all 
the lepton-scalar couplings can be small if all vacuum expectation values 
(VEVs),  except  the one controlling the 
$SU(3)$ breaking, are of the order of the electroweak scale 
and, if there are no flavor changing neutral currents (FCNC) in the leptonic 
sector. Both conditions may not be natural in the minimal version 
of the model. The former because the sextet is introduced only to give
mass to the leptons so its VEV may be of the order of a few GeV.
The later because if we want to avoid FCNC in the lepton sector it is
necessary to impose a discrete symmetry which does not belong to the minimal 
331 model. In the model the $SU(3)_L\otimes U(1)_N$ triplets 
$\eta=(\eta^0\,\eta^-_1\,\eta^+_2)^T\sim({\bf3},0)$ and 
$\rho=(\rho^+\,\rho^0\,\rho^{++})^T\sim({\bf3},+1)$ give mass to the quarks. 
(The third triplet $\chi\sim(\chi^-\,\chi^{--}\,\chi^0)^T$ is out of our 
concern here.)
If the first family transforms in a different way from the other two, 
the quark $u$ mass is given by the VEV of the $\eta$, here denoted by 
$v_\eta$; if the third family is which transforms differently, it is the 
quark $t$ which gets its mass from $v_\eta$. However, since the mixing matrix 
of the charge 2/3 quarks is not trivial the general case interpolates between 
these two cases. Hence, the vacuum expectation values $v_\eta$ and $v_\rho$ 
are of the order of the electroweak scale, {\it i.e.}, $v^2_\eta+v^2_\rho=
(246\,\mbox{GeV})^2$.
As we said before, the scalar sextet $S\sim({\bf6},0)$ 
\begin{equation}
 S=\left( 
\begin{array}{ccc}
\sigma _1^0 & \frac{h_2^{-}}{\sqrt2} & \frac{h_1^{+}}{\sqrt2} \\ 
\frac{h_2^{-}}{\sqrt2} & H_1^{--} & \frac{\sigma _2^0}{\sqrt2} \\ 
\frac{h_1^{+}}{\sqrt2} & \frac{\sigma _2^0}{\sqrt2} & H_2^{++}
\end{array}
\right)
\label{es}
\end{equation}
is necessary in order to give to the charged leptons an arbitrary mass.
We will denote $\langle \sigma^0_2\rangle=v_S$ the VEV of $\sigma_2^0$ the 
neutral component of the sextet. 
The other one, $\sigma^0_1$ does not
gain a nonzero VEV if the neutrinos must remain massless.

The Yukawa couplings in the lepton sector are 
\begin{equation}
-{\cal L}_{l}=\frac{G_{ab}}{\sqrt2}\,
\overline{\psi_{aiL}}\,(\psi_{bjL})^cS_{ij}+
\frac{1}{2}\epsilon_{ijk} G'_{ab}\overline{\psi_{aiL}}\,(\psi_{bjL})^c\eta_k
+H.c.,
\label{yuka}
\end{equation}
where $\psi=(\nu \,l \,l^c)^T$ and $a,b=e,\mu,\tau$; $i,j$ are $SU(3)$ 
indices; $G_{ab}$ and $G'_{ab}$ are symmetric and anti-symmetric complex 
matrices, respectively. (The model can have $CP$ violation in the leptonic 
sector~\cite{liung}.) 

We stress once more that this is the minimal 331 model because if we want to 
avoid in Eq.~(\ref{yuka}) the coupling with the triplet, since only the sextet
is necessary for giving all charged leptons a mass, we have to impose a 
discrete symmetry.
(Only in this case there is not FCNC in the lepton sector.)
Hence, the mass matrix of the charged leptons has the form 
\begin{equation}
M_l= \frac{1}{\sqrt2}\left(G_{ab}v_S+G'_{ab}v_\eta\right),
\label{ml}
\end{equation}
and it is diagonalized by the bi-unitary transformation 
$E^\dagger_L M_lE_R={\rm diag}(m_e,m_\mu,m_\tau)$ with $E_L$ and $E_R$ defined
in Eq.~(\ref{1}). However, the biunitary transformation does not diagonalize
$G$ and $G'$ separately. 
Thus, we have FCNC and there are Yukawa couplings which are 
not proportional to the lepton masses. 

The model has four singly charged and two doubly charged physical scalars,
four $CP$-even and two $CP$-odd neutral scalars.
Let us consider the doubly charged and neutral scalar Yukawa interactions
with the sextet in Eq.~(\ref{es}).
$H^{++}_1$ is a part of a complex triplet under $SU(2)_L\times U(1)_Y$ with its
neutral partner having vanishing vacuum expectation value (if neutrinos do not
get Majorana masses). There are also a doubly charged $H^{++}_2$ which is a
singlet of $SU(2)_L$ and the neutral Higgs $\sigma^0_2$ which is part of a 
doublet of $SU(2)_L$. Hence we have the respective Yukawa interactions
proportional to
\begin{equation}
\overline{l_L} {\cal K}_{LL} l_L^c H^{--}_1+
\overline{l^c_L} {\cal K}_{RR}l_{R}H^{++}_2 
+[\overline{l_L}\;{\cal K}_{LR}l_R+\overline{l_L^c}{\cal K}^T_{LR}l^c_L]\,
\sigma^0_2+H.c.,
\label{2p}
\end{equation}
where we have denoted ${\cal K}_{LL}= E^T_LGE^*_L$; 
${\cal K}_{RR}= E^T_R GE_R$ and ${\cal K}_{LR}=E^\dagger_LGE_R$. These
matrices are unitary only when $G$ is real.

We see that since the unitary matrices $E_{L,R}$ diagonalize 
$M_l$ in Eq.~(\ref{ml}), ${\cal K}_{LL},\overline{K}_{RL}$ and ${\cal K}_{RL}$
are not diagonal matrices, thus their matrix elements are arbitrary and only 
constrained by perturbation theory,  by their contributions to the charged 
lepton masses and by some purely leptonic processes. 

\begin{figure}[ht]
\centering\leavevmode
\epsfxsize=350pt 
\epsfbox{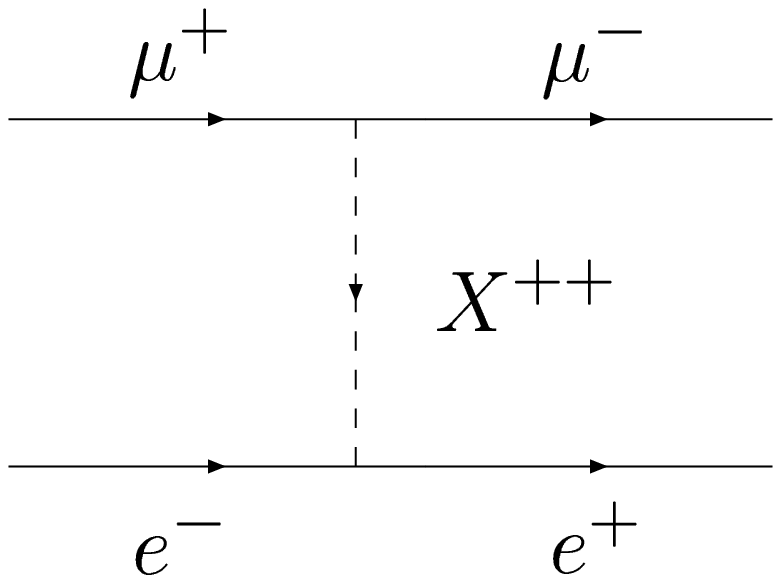}
\vglue -0.009cm
\caption{Contribution to the $M\to\overline{M}$ conversion. $X^{++}$ denotes a 
vector $U^{++}_\mu$ or a scalar bilepton $H^{++}_{1,2}$.}
\label{fig1}
\end{figure}

\begin{figure}[ht]
\centering\leavevmode
\epsfxsize=350pt 
\epsfbox{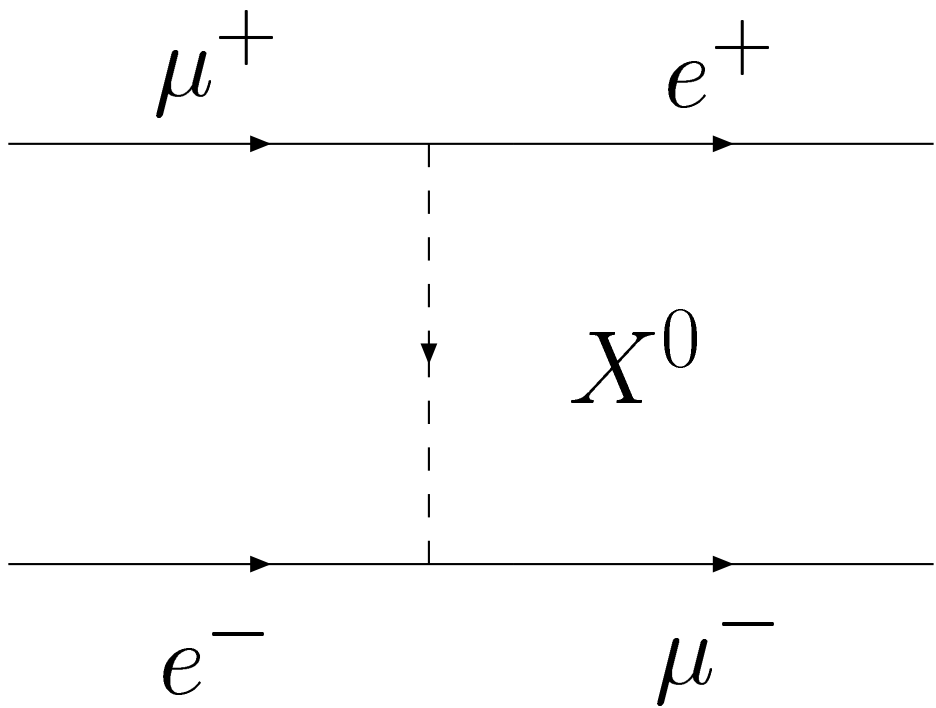}
\vglue -0.0009cm
\caption{Contribution to the $M\to\overline{M}$ conversion. $X^{0}$ denotes
a neutral scalar or pseudoscalar.}
\label{fig2}
\end{figure}

As we said before, since there 
are already two scalar triplets which give the appropriate mass to the 
$W^\pm$ and $Z^0$ vector bosons, it is not necessary the $v_S$ be of the 
same order of magnitude than the other vacuum expectation values which are 
present in the model, $v_\eta$ and $v_\rho$. 
For instance it is possible that $v_S\approx 10$ GeV and 
$\vert G\vert\approx1$. In this case, the contributions of the doubly charged 
scalars to the muonium-antimuonium transition  can be as important as the 
contribution of the vector bilepton. There are also new contributions 
involving FCNC through the neutral scalar exchange,
can also give important contributions to the $M\to\overline{M}$ conversion as 
it has been suggested in Ref.~\cite{hw}.
We see that in the 331 model all contributions shown in Figs.~\ref{fig1} and 
\ref{fig2} do exist.
Since there are several contributions to the $M\to\overline{M}$ conversion
it is still possible to have some cancellations among the scalar and vector
bilepton contributions.

In order to appreciate a little bit more the muonium-antimuonium transition in
the 331 model let us give a brief review of the theoretical results so far 
known.
Many years ago, Feinberg and Weinberg~\cite{fw} used a $(V-A)^2$ Hamiltonian
with the four-fermion effective coupling equal to the usual
$\beta$-decay coupling constant $C_V$, in order to study the 
$M\to\overline{M}$ conversion. Let us here denote it by $G_{M\bar{M}}$.
The transition amplitude is proportional to
$\delta=16G_{M\bar{M}}/\sqrt{2}\pi a^3$ where $a$ is the Bohr radius.
More  recently, the same transition was studied in the context of models with 
doubly charged Higgs; in this case the effective Hamiltonian is of the 
$(V\pm A)^2$ form~\cite{ah,mls,ck}. In this case $G_{M\bar{M}}$ is given by 
the product of two Yukawa couplings~\cite{mls,ck}, so in all these cases the 
sing of the effective coupling $G_{M\bar{M}}$ is undetermined.  
On the other hand, in models with doubly charged vector bilepton the respective
Hamiltonian is of the $(V-A)\times(V+A)$ form~\cite{teo} with a  four-fermion 
effective coupling given by $G_{M\bar{M}}/\sqrt{2}=-g^2/8M^2_U$, being $M_U$ 
the vector bilepton mass and $g$ the $SU(3)$ coupling constant. 
Hence, in this case always $G_{M\bar{M}}<0$. 

On the other hand, in $(V\pm A)^2$ models the transition amplitude
is the same for the singlet and triplet muonium given above~\cite{fw,ah,mls} 
but in $(V-A)\times(V+A)$ models we have 
$\delta=-8G_{M\bar{M}}/\sqrt{2}\pi a^3$ for the triplet muonium state and 
$\delta=24G_{M\bar{M}}/\sqrt{2}\pi a^3$ for 
the singlet state~\cite{teo}. 

In the 331 model there are also neutral scalars and pseudoscalars which, as we 
said before, have flavor changing neutral interactions in the lepton sector.
It has been shown that pseudoscalars do not induce conversion for triplet 
muonium, while both pseudoscalars and scalars contribute for the singlet 
muonium. 
We see that it is in fact possible a 
cancellation among the contributions to the $M\to\overline{M}$ transition due 
to scalars and those due to doubly charged vector. It means that
separate measurements of singlet vs. triplet $M\to\overline{M}$ conversion 
probabilities can distinguish among neutral scalar, pseudoscalar and doubly
charged Higgs induced transition~\cite{hw}. 
Such measurements can also distinguish doubly charged vector bileptons from 
scalar contributions. 
 
The $M\to\overline{M}$ transition also can be measured in matter.
In this case the collisions make the amplitude add incoherently~\cite{fw}.
However in matter the conversion is strongly suppressed mainly due to
the loss of symmetry between $M$ and $\bar{M}$ due to the possibility of
$\mu^-$ transfer collisions involving $\bar{M}$~\cite{fw,vh2,kj}. 
Hence, all those data together when 
available will allow to constrain models with several sort of fields inducing 
the muonium--antimuonium transition.

If all these effects are present or not in the 331 model depend on the value 
of the parameters. We have argue above that this may be the fact since there 
are flavor changing neutral interactions in the Higgs-lepton sector and also 
one of the vacuum expectation values may be of the order of some GeVs.    

\begin{figure}[ht]
\centering\leavevmode
\epsfxsize=\hsize
\epsfbox{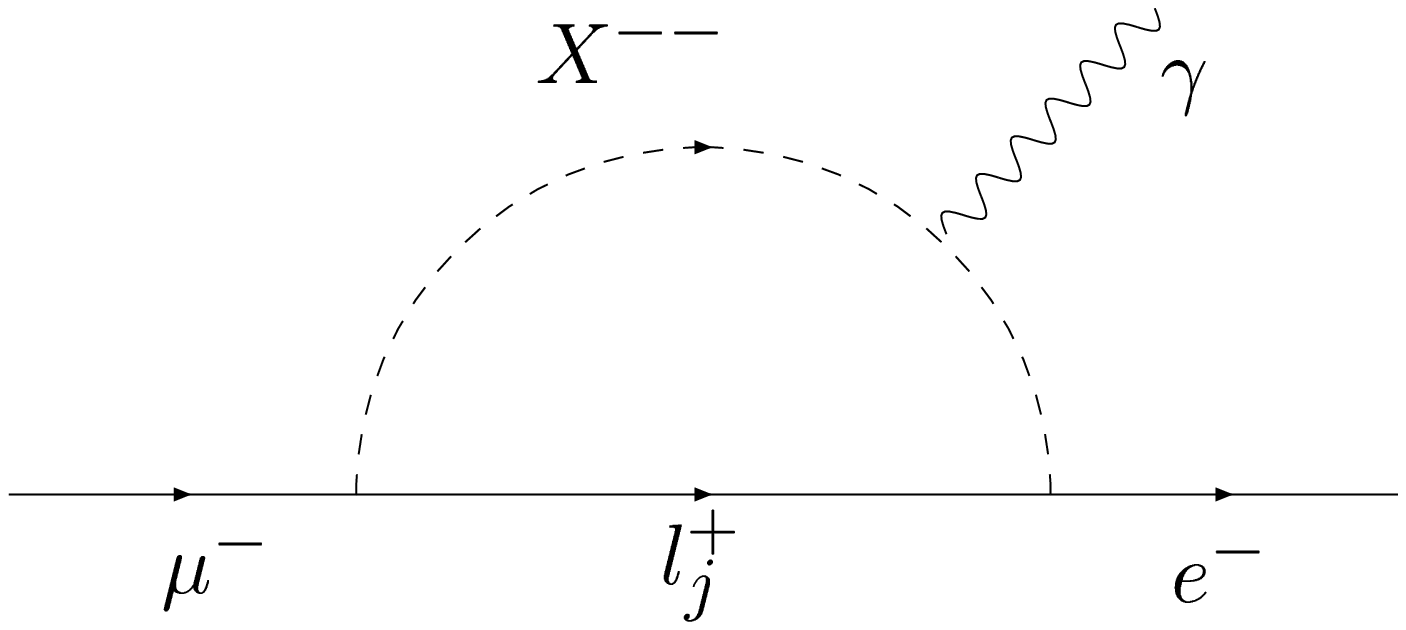}
\vglue -0.0009cm
\caption{Contribution to $\mu\to e\gamma$. As in Fig.~1 $X^{--}$ denotes
any of the doubly charged bosons.}
\label{fig3}
\end{figure}

It is usually considered that the $\mu\to e\gamma$ decay 
imposes stronger constraints on a given model than the $M\to\overline{M}$ 
transition.
So, some model builders consider situations in which $\mu\to e\gamma$ 
is forbidden by a discrete symmetry~\cite{ck}. However in the 331 model 
the interactions which induce the $\mu\to e\gamma$ decay are  
$\overline{\nu_L}\,{\cal K}_{LR} l_Rh^+_1$,
$\overline{\nu_L}\,{\cal K}^T_{LL}l_L^c h^-_2$,  
$[ \overline{\nu_L} {\cal K}^{\prime}_{LR}l_R+
\overline{(l^c)_L} {\cal K}^{\prime T}_{LR}(\nu_L)^c]\eta^-_1$ and
$[\overline{\nu_L} {\cal K}^{\prime}_{LL}(l_L)^c-
\overline{l_L}{\cal K}^{\prime}_{LL}(\nu_L)^c]\eta^+_2$,
with ${\cal K}^{\prime}_{LR}=E^\dagger_LG^\prime E_R$ and
$K^\prime_{LL}=E^T_LG^\prime E^*_L$. The decay
$\mu \to e\gamma$ as shown in Fig.~\ref{fig3} has contributions of
the vector $U^{--}$ and scalars $H^{--}_{1,2}$ bileptons. The interactions
in Eqs.~(\ref{bu}) and (\ref{2p}) involve different mixing matrices, hence,
if all bosons have masses of the same order of magnitude, as it is in fact 
expected in the 331 model (see below), we can have some cancellations among all
the contributions. Notice that those matrices are unitary only when $G'$ is 
real, like the matrices in Eq.~(\ref{2p}).

Notice also that the $\mu\to e\gamma$ decay is dominated by the lepton $\tau$ 
contributions, thus it implies strong constraints on the mixing angles 
involving this lepton.  

\begin{figure}[ht]
\centering\leavevmode
\epsfxsize=350pt 
\epsfbox{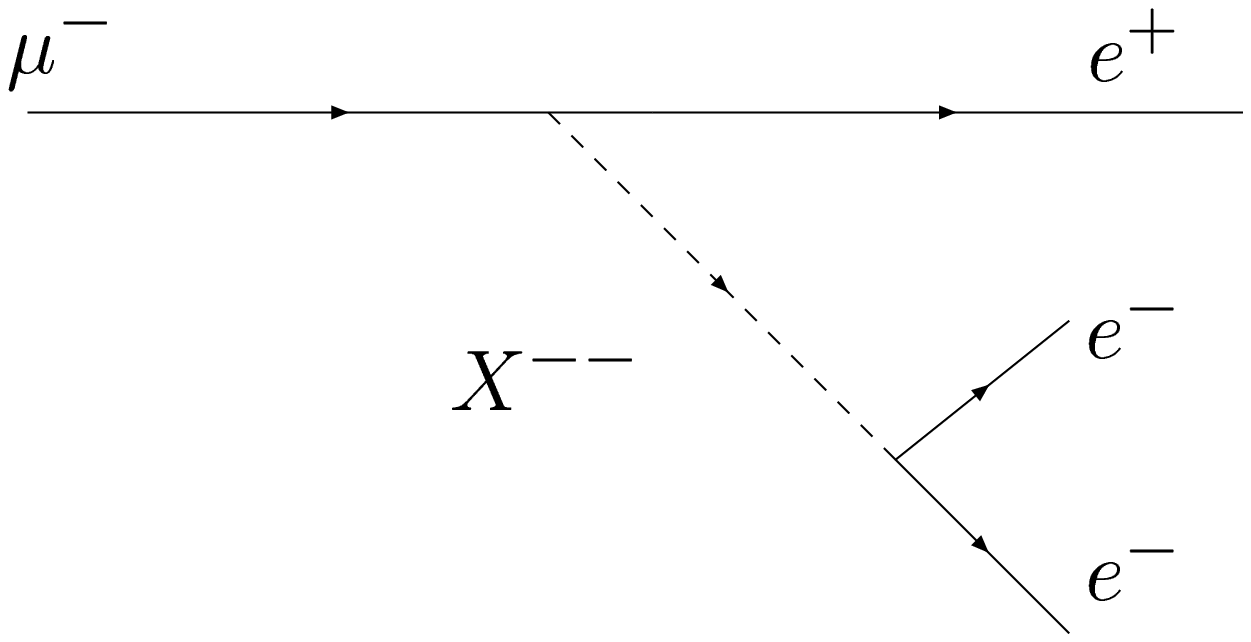}
\vglue -0.0009cm
\caption{Contribution to $\mu\to eee$. As in Fig.~1 $X^{--}$ denotes
any of the doubly charged bosons.}
\label{fig4}
\end{figure}

Another potential trouble for the model is the $\mu\to eee$ decay shown in 
Fig.~\ref{fig4}.

Here the amplitudes of the exchange of $U^{--}$, $H^{--}_1$ and $H^{--}_2$ are
proportional to ${\cal K}_{\mu e}{\cal K}_{ee}$
$({\cal K}_{LL})_{\mu e}({\cal K}_{RL})_{ee}$ and 
$({\cal K}_{RL})_{\mu e}({\cal K}_{RL})_{ee}$ respectively, as it can be seen 
from Eqs.~(\ref{bu}) and (\ref{2p}). Thus in this case again a cancellation 
among the contributions may occur if all bosons have masses of the same order 
of magnitude, or it may be suppressed by the $\mu e$ matrix element.
 
Summarizing, the bound of Ref.~\cite{exp} is applied only for a range of the 
parameters in the model and if the sextet is the only Higgs which 
couples to leptons. In this case the neutral currents given in Eq.~(\ref{2p}) 
are diagonal and there is no FCNC in the lepton sector and all the 
sextet-lepton couplings are proportional to 
${\cal K}_{RL}=\sqrt{2}m_l/v_S$ where $m_l$ is the lepton mass. 
If $v_S$ is of the order of 100 GeV the main
contribution to the $M\to\overline{M}$ conversion comes from the interaction 
in Eq.~(\ref{bu}) and it constrains the mass of the $U$-vector boson and the 
mixing angles of the matrix ${\cal K}$ as discussed early.
Hence in the minimal 331 model the contributions in Fig.~1 (a), 1(c) 
and 1(d) of Ref.~\cite{exp}, here summarized in Figs.~\ref{fig1} and 
\ref{fig2}, do exist and its experimental data do not, in straightforward way, 
apply to the model.

Finally we would like to remark that although the model predicts that there 
will be a Landau pole at the energy scale $\mu$ when 
$\sin^2\theta_W(\mu)=1/4$, it is 
not clear at all what is the value of $\mu$. In fact, it has been argued that 
the upper limit on the vector bilepton masses is 3.5 TeV~\cite{jj}. 
Any way the important thing is that in this model the ``hierarchy 
problem'' {\it i.e.}, the existence of quite different mass scales, is less 
severe than in the standard model and its extensions since
no arbitrary mass scale (say, the Planck scale) can be introduced in the
model. In particular, it is a very well known fact that the masses
of fundamental scalars are sensitive to the mass of the heaviest particles
which couple directly or indirectly with them. Since in the 331 model the 
heaviest mass scale is of the order of a few TeVs there is not a ``hierarchy
problem'' at all. 
This feature remains valid when we introduce
supersymmetry in the model. Thus, the breaking of the supersymmetry is also
naturally at the TeV scale in this 331 model. 

The $M\to\overline{M}$ transition deserves indeed more experimental studies. 
On the other hand, the matrices $G$ and $G^\prime$ can be complex, we can have
CP violation in the present model~\cite{liung}. Hence,
experimental difficulties apart, this system in vacuum could be useful for
studying $CP$ and $T$ invariance in the lepton sector: 
by comparing $M\to\overline{M}$
with $\overline{M}\to M$ transitions as it has been done  recently in the
$K^0\to \bar{K}^0$ and  $\bar{K}^0\to {K}^0$ case~\cite{ktev}.
 
\acknowledgments 
This work was supported by Funda\c{c}\~ao de Amparo \`a Pesquisa
do Estado de S\~ao Paulo (FAPESP), Conselho Nacional de 
Ci\^encia e Tecnologia (CNPq) and by Programa de Apoio a
N\'ucleos de Excel\^encia (PRONEX).

\vglue 0.01cm

\newpage
\vglue 0.01cm

\newpage

\vglue 0.01cm

\newpage

\vglue 0.01cm
\end{document}